\documentclass[a4paper]{article}
\usepackage[latin1]{inputenc} % ou \usepackage[utf8]{inputenc}
\usepackage[T1]{fontenc} % ou \usepackage[OT1]{fontenc}
\usepackage{RR,RRthemes}
\usepackage{hyperref}
\graphicspath{{./logos/}}
\RRdate{November 2011}
\RRauthor{% les auteurs
Ichrak Amdouni %\thanks{Footnote for first author}%
  \and
Pascale Minet %\thanks{Footnote for second author}%
 \and
Cedric Adjih
}
\authorhead{Amdouni \& Minet \& Adjih}
\RRtitle{\OSERENA{}, Un algorithme de coloriage optimis\'e pour les r\'eseaux sans fil denses ou large \'echelle}

\RRetitle{\OSERENA{}, an Optimized Coloring Algorithm for Dense or Large Scale Wireless Networks}
\titlehead{\OSERENA{}, an optimized graph coloring algorithm}
\RRresume{Le but de ce rapport de recherche est de pr\'esenter \OSERENA{}\NoteBlinded{} "Optimized SchEduling RoutEr Node Activity", un algorithme de coloriage distribu\'e optimis\'e pour les r\'eseaux sans fil denses. Avec cette optimisation, la densit\'e du r\'eseau a un impact extr\^emement r\'eduit sur la taille des messages \'echang\'es pour colorier le r\'eseau. Par ailleurs, le nombre de couleurs utilis\'e pour colorer le r\'eseau n'est pas affect\'e par cette optimisation. %Nous le temps de convergence évalué que le nombre de tours n'est pas impacté par cette optimisation.
Nous d\'ecrivons dans ce rapport de recherche les propri\'et\'es de l'algorithme et prouvons sa correction et sa terminaison. Les r\'esultats des simulations mettent en \'evidence les gains consid\'erables en bande passante.
}
\RRabstract{
The goal of this research report is to present \OSERENA{}\NoteBlinded{} "Optimized SchEduling RoutEr Node Activity", a distributed coloring algorithm optimized for dense wireless networks. Network density has an extremely reduced impact on the size of the messages exchanged to color the network. Furthermore, the number of colors used to color the network is not impacted by this optimization. %We The convergence time evaluated as the number of rounds is not impacted by this optimization. 
We describe in this research report the properties of the algorithm and prove its correctness and termination. Simulation results point out the considerable gains in bandwidth. 
}
\RRmotcle{OSERENA, coloriage distribu\'e, optimisation, r\'eseaux sans fil denses, r\'eseaux large \'echelle, surco\^ut r\'eduit, nombre de couleurs, taille des messages, temps de convergence...}
\RRkeyword{OSERENA, distributed coloring, optimization, dense wireless networks, large scale networks, reduced overhead, number of colors, message size, convergence time...}
\RRprojet{Hipercom}  % cas d'un seul projet
\RRdomaine{3} % cas du domaine numero 1
\RRthemeProj{hipercom} % theme du projet Hipercom
\RRdomaineProjBis{hipercom} % domaine du projet Hipercom
\RCParis % Paris Rocquencourt
\newcommand{\BlindReview}[2]{#2} % <-------- NORMAL VERSION

\newcommand{\OSERENA}{OSERENA} % <----- Not Blinded
\newcommand{\SERENA}{SERENA}   % <----- Not Blinded
\newcommand{\NoteBlinded}{}

\newcommand{\figsize}{4.5in}

\usepackage{subfigure}
\usepackage{color}%ichrak
\usepackage{graphicx}
\usepackage{mathenv}
\usepackage{amssymb}%cedric

\def\QED{\mbox{\rule[0pt]{1.5ex}{1.5ex}}} %Pascale
\def\proof{\hspace*{-0.8cm}{{\itshape Proof: }}} %Pascale
\def\endproof{\hspace*{\fill}~\QED\par\endtrivlist\unskip} %Pascale
\newtheorem{lemma}{Lemma} %Pascale
\newtheorem{property}{Property} %Pascale
\begin{document}
\RRNo{7785}
\makeRR   % cas d'un rapport de recherche
%% \makeRT % cas d'un rapport technique.
%% a partir d'ici, chacun fait comme il le souhaite
\tableofcontents
\newpage

%\title{\OSERENA{}: a Coloring Algorithm Optimized for Dense Wireless Networks}


%----------

%----------


% \fi

%\pagestyle{plain} % Pascale for paper submission only

\section{Introduction and state of the art \label{Intro}}
%========================================================
Graph coloring can be seen as a specific case of graph labeling: labels, usually called colors, are assigned to vertices (respectively edges) of a graph subject to certain constraints. Depending on which graph element is colored, we obtain vertex or edge coloring. For both, the objective is to minimize the number of colors needed to color the whole graph. Typically, the constraint considered for $h$-hop vertex coloring, with $h$ an integer $\geq 1$ is: no two vertices that are $k$-hop neighbors with $1 \leq k \leq h$ have the same color. The Vizing's theorem \cite{Viz64} states that the minimum number of colors needed to 1-hop color a graph, number denoted $\chi$, meets $\Delta \leq \chi \leq \Delta +1$, where $\Delta$ is the maximum degree of the graph. The $h$-hop node coloring problem has been proved NP-complete in \cite{Gar79} for $h=1$ and 
%------------------------------
\BlindReview{
in \cite{C83} for any $h>1$. 
}{
in \cite{C83,RR11} for any $h>1$. 
}
%------------------------------
This explains why heuristics are used for large graphs. Authors of \cite{HDD03} compare the performances of different heuristics for edge coloring over standard benchmarks (taken from a list of 119 graphs given at CP2002) for small graphs ($<500$ nodes). They show that among the tested heuristics, $ant1$ often finds the optimum or a number of colors differing not much. A survey of local search methods ($TabuCOL$, simulated annealing, neighborhood search, and clustering guided search) can be found in \cite{GH06} and \cite{WH11}. If these algorithms are efficient for small graphs, it is no longer the case with large random graphs \cite{WH11}. Hybrid algorithms can be used, as well as the extraction of large independent sets from the graph to obtain a smaller residual graph easier to color (see for instance $EXTRACOL$ that needs 2.5 hours to color 1000 nodes with a density of 5 in \cite{WH11}). The main performance criteria of a coloring algorithm are the number of colors and the time needed to color the considered graph. Of crucial interest is the approximation ratio of coloring algorithms that is defined as the ratio of the number of colors obtained by the algorithm to the optimal number. A well-known coloring algorithm is FirstFit \cite{FirstFit} that sequentially assigns colors to nodes. Each node is colored with the first available color. Depending on the coloring order, different coloring results are obtained. Approximation ratio of coloring algorithms
for grids, triangular lattices and hexagonal graphs can be found in \cite{BHH10}. All nodes having the same color constitute a class. Hence, graph coloring can also be seen as determining independent sets of maximum size.\\

Coloring has been applied to wireless networks to improve medium access efficiency. Thus, with node coloring, nodes access the medium in time slots corresponding to their color \cite{rajendran03, ZMAC, iwcmc08}. Only nodes that do not interfere can transmit simultaneously, hence collisions are avoided while spatial reuse of the bandwidth is provided. The smaller the number of colors, the shorter the activity period in data gathering applications \cite{rajendran05, FlexiTP, TDMA-ASAP}. A color can be mapped into a channel, that is why graph coloring has been applied to channel assignment reducing radio interferences \cite{HC11}.\\

Running a distributed coloring algorithm on WSNs (Wireless Sensor Networks) is very challenging because of their strong limitations. They have low capacity of storage and computing, low energy especially for battery operated nodes and the network bandwidth is also limited. That is why algorithms supported by WSNs must be of low complexity. More challenging are dense WSNs, where a node cannot maintain its 2-hop neighbors because of memory limitation and a single message cannot contain all the information relative to the 2-hop neighbors of a node. Examples of dense WSNs are given by smart dust where microelectomechanical systems called MEMS can measure temperature, vibration or luminosity. Applications can be  monitoring of building temperature, detection of seismic events, monitoring of pollution, weather prediction for vineyard protection... In this research report, we show how to optimize a coloring algorithm for dense WSNs. We present \OSERENA{}, an optimized version of the node coloring algorithm \SERENA{}~\cite{FutureInternet2}. The optimization consists in the reduction of the algorithm overhead in both sizes of data stored and messages exchanged to color the network. Indeed, \OSERENA{} does not require neither the storage nor the exchange of neighbors up to two hops. Furthermore, we prove that \OSERENA{} keeps the same number of colors as \SERENA{}. Moreover, \OSERENA{} produces a small convergence time that is equal, most of the time, to the time needed by \SERENA{} to color the algorithm.\\

The research report is organized as follows. In Section~\ref{Optimization}, we present \OSERENA{}, a 3-hop node coloring algorithm that is optimized for dense networks. %We give its principles and prove its properties.
In section~\ref{sec:PropOserena}, we present the properties of \OSERENA{} regarding the correctness, the overhead induced, and its convergence time. We also prove that \OSERENA{} is equivalent to a centralized version of 3-hop node coloring.
In Section~\ref{PerfSimul}, we evaluate the performance of \OSERENA{} for many network configurations, varying network size and node density. We show that unlike the previous work in \cite{FutureInternet2}, \OSERENA{} keeps a number of rounds similar to the number of rounds induced by \SERENA{} (the unoptimized version) to color the network, while using smaller messages. This property is illustrated through an extensive performance evaluation by means of simulations.
%The validity of our simulation tool is proved in Section \ref{PerfMeasure} by confrontation with measures obtained on SensLAB \cite{Senslab}, a real WSN platform on which \OSERENA{} is implemented as well as a Neigborhood Discovery protocol. Finally, we conclude in Section \ref{Conclusion}.

\section{Coloring optimized for dense networks \label{Optimization}}
%======================================================================
The goal of this section is to make possible the use of the coloring algorithm in dense wireless sensor networks. We show how to reduce the overhead in terms of 1) memory required to store the data maintained by each node and 2) bandwidth used by exchanging messages between neighbors. Of course this overhead reduction must not decrease the performance of the coloring algorithm: the number of colors and the time needed to color all network nodes must be kept small. First, we give some definitions followed by the basic principles of 3-hop node coloring.

\subsection{Assumptions and definitions}
%-----------------------------------------

\subsubsection{Central assumptions and definitions}
~\\
The type of node coloring needed to support a given application depends on the type of:
\begin{itemize}
\item \textit{communications supported}: unicast and/or broadcast;
\item \textit{application}: general where any node is likely to exchange information with any neighbor node or on the contrary tree type where a node exchanges information only with its parent and its children in the data gathering tree;
\item \textit{acknowledgement for unicast transmissions}: immediate or deferred.
\end{itemize}
In this report, we focus on 3-hop node coloring, which was proved in~\cite{FutureInternet2} to be necessary to support general communications, where unicast transmissions are immediately acknowledged. We assume an ideal environment where:\\
{\em Assumption A0:} All links are symmetric and stable.\\
{\em Assumption A1:} Each node has a unique address in the network.\\
{\em Assumption A2:} Any node does not prevent the correct receipt of any other node out of its transmission range.\\

A 3-hop node coloring is said {\em valid} if and only if no two 1-hop, 2-hop or 3-hop nodes have the same color. The smaller the number of colors obtained, the better the coloring algorithm.\\

The time complexity of a coloring algorithm is generally evaluated in terms of rounds. By definition, a {\em round} is such that any node receives the messages sent by its 1-hop neighbors, processes them and broadcasts its own message to its 1-hop neighbors. The space complexity is given by the number and size of messages sent per node.

\subsubsection{Further assumptions and simplifications}
\label{sec:simplify}
~\\
In some sections, we will assume a more specific model, 
closely related to a common model for wireless sensor networks: 
the unit disk graph model \cite{bib:CCJ02}. Hence:
\begin{itemize}
\item Nodes are modeled as a set of points in the 2-dimensional plane.
\item A uniform transmission range $R$ is defined.
\item A node receives a packet from another node, if and only if,
   its distance is lower than $R$.
\item There are no losses.
\end{itemize}

\noindent 
The same model is applied for instance for simulations in section \ref{PerfSimul}.\\
Furthermore, in some calculation, we also make the following
approximation: \\
{\em Assumption (approximation) A3:}  
we equate distance to number of hops (e.g. a node at distance
between $R$ and $2 R$ from another node, is assumed to be at $2$ hops).\\
The assumption is valid asymptotically when the density converge towards
infinity ; for a more detailed exploration of the exact relationship
between number of hops and distance, see for instance \cite{XGA07}.

\subsection{Basic principles of 3-hop node coloring}
%---------------------------------------------------
\noindent In \SERENA{}, any node $u$ proceeds as follows to color itself:
\begin{enumerate}
\item Node $u$ characterizes the set $\mathcal{N}(u)$ of nodes that cannot have the same color as itself. The set $\mathcal{N}(u)$ is the set of neighbors up to 3-hop from $u$ in 3-hop node coloring.

\item Node $u$ computes its priority, denoted $priority(u)$. This priority consists of two components: the most important one is denoted $prio(u)$. It can be equal to the number of nodes up to 2-hop (resp. 3-hop) from $u$. We will see later the exact value taken in \OSERENA{}. The second component of $priority(u)$ denotes the $address$ of the node. By definition, node $u$ is said to have a priority higher than node $v$ if and only if:
\begin{itemize}
\item either $prio(u) > prio(v)$;
\item or $prio(u)= prio(v)$\\
\hspace*{10pt} and $address(u) < address(v)$.
\end{itemize}

\item Node $u$ applies the two following rules:
\begin{itemize}
\item \textbf{Rule R1}: Node $u$ colors itself if and only if it has a priority strictly higher than any uncolored node in $\mathcal{N}(u)$.
\item \textbf{Rule R2}: To color itself, node $u$ takes the smallest color unused in $\mathcal{N}(u)$.
\end{itemize}
\end{enumerate}

\subsection{Motivations and optimization principles}
%---------------------------------------------------
\noindent This distributed coloring algorithm proceeds by iterations or rounds, where nodes exchange their $Color$ message. In its simplest implementation, the $Color$ message would include the address, the priority and the color of 1) the node $u$ itself, 2) its 1-hop neighbors in $\mathcal{N}(u)$, as well as 3) its 2-hop neighbors in $\mathcal{N}(u)$. The data locally maintained by any wireless sensor would include these data as well as the priority and color of any neighbor up to $3$-hop. It is well known that the average number of nodes in the neighborhood up to 2-hop is equal to $4 \cdot density$, where $density$ stands for the average number of nodes in the disk of radius $R$, where $R$ is the transmission range. Such an overhead can be unacceptable for wireless sensors with limited storage and processing capabilities as well as low residual energy. Dense networks with limited bandwidth, low energy and a short MAC frame size become challenging for a coloring algorithm.
That is why, we propose in this report an optimization of the coloring algorithm reducing the size of $Color$ messages exchanged and the size of data structures maintained, while keeping a low complexity. We also show that this overhead reduction does not increase the convergence time of the coloring algorithm.
\noindent The optimization principles are based on the following remarks:
\begin{itemize}
\item It is necessary that any node $u$ knows the highest priority taken by its uncolored neighbors up to 3-hop in order to apply Rule R1. Furthermore, node $u$ must send information concerning itself, its 1-hop and 2-hop neighbors to let its one-hop neighbors know information about their 1-hop, 2-hop and 3-hop neighbors.
Hence, node $u$ must send its priority, the highest priority taken by its uncolored 1-hop neighbors as well as the highest priority taken by its uncolored 2-hop neighbors. However, sending only one highest priority of the uncolored 1-hop or 2-hop neighbors would delay the coloring since the information update will be slow.
%To better understand this constraint, let us consider three nodes $u$ $v$ and $w$ where $u$ is 1-hop neighbor of $v$, itself 1-hop neighbor of $w$ and assume $prio(v)\l prio(u)\l prio(w)$. When $w$ color itself at round $r$,
This would not suffice to color any wireless network with the same number of rounds as \SERENA{}.
Indeed, node $v$, 2-hop away from node $u$ colored at round $r$ would not know at round $r+2$ that it has the highest priority. Hence, MORE THAN ONE highest priority at respectively 1-hop and 2-hop must be maintained and sent, unlike the version briefly presented in \cite{FutureInternet2}. We will wee in Section~\ref{subsec:ReducedOverhead} how to compute the near optimal number of priorities to maintain at one-hop and two-hop respectively. Notice that the highest priority at 3-hop is locally computed and not sent.
\item Similarly for the color, node $u$ must know the colors already used in its neighborhood up to 3-hop. However, it does not matter $u$ to know which node up to 3-hop has which color, but only which colors are taken at 1-hop, 2-hop and 3-hop respectively. That is why, we use the fields $color\_bitmap1$, $color\_bitmap2$ and $color\_bitmap3$ for the bitmaps of colors used at 1-hop, 2-hop and 3-hop respectively.
\end{itemize}

%\newpage
\subsection{\OSERENA{}: Optimized coloring algorithm}
%----------------------------------------
\subsubsection{The $Color$ message \label{subsubsec:ColorMsg}}
From simulation feedback, we have noticed that the assignment:\\
$prio(u)= $number of neighbors up to 2 hops\\
outperforms the assignment:\\
$prio(u)= $number of neighbors up to 3 hops from $u$, or a random assignment. \\
However, as \OSERENA{} avoids the expensive computation of the list of neighbors up to 2 hops, \OSERENA{} defines for any node $u$, $prio(u)$ as the number of its neighbors + the sum of the number of 1-hop neighbors of its 1-hop neighbors. This computation is done during the initialization of the coloring algorithm. We also define $max\_prio1(u)$ as:
\begin{itemize}
\item the four highest priorities of the uncolored 1-hop neighbors of $u$, if four such nodes exist;
\item the priority of the only three (respectively two, respectively one) uncolored 1-hop neighbor, if only three (respectively two, respectively one) such nodes exist;
\item empty, denoted $\varnothing$, if none exists.\\
\end{itemize}
\vspace{-7pt}
We then have the following notation:\\
$max\_prio1(u)= Max4_{\ v\ uncolored \in 1hop(u)\ }priority(v)$.\\

\noindent Similarly, we define $max\_prio2(u)$ as the three highest priorities of the uncolored 1-hop neighbors of the 1-hop neighbors of $u$, if they exist. We then have:\\
$max\_prio2(u)= Max3_{\ v \in 1hop(u)\ } max\_prio1(v)$.\\

\noindent The variable $max\_prio3(u)$ is defined as the highest priority of the uncolored 1-hop neighbors of the 1-hop neighbors of the 1-hop neighbors of $u$. We get:\\
$max\_prio3(u)= Max_{\ v \in 1hop(u)\ } max\_prio2(v)$.\\

\noindent The computation of $max\_prio1(u)$, $max\_prio2(u)$ and $max\_prio3(u)$ is done from the $Color$ messages received during the current round. The values computed for  $max\_prio1(u)$ and $max\_prio2(u)$ are inserted in the $Color$ message sent by node $u$.\\

\noindent It follows that the $Color$ message sent by any node $u$ contains $priority(u)$, $max\_prio1(u)$ and $max\_prio2(u)$, as well as the color of $u$, the bitmap of colors used at 1-hop from $u$, denoted $bitmap1(u)$ and the bitmap of colors used at 2-hop from $u$, denoted $bitmap2(u)$.\\ 
% Notice that the size of the $Color$ message is variable for two reasons. Since $max\_prio1$ can contain 0, 1, 2, 3 or 4 priority values, its size is given in the field $size\_max\_prio1$. Idem for $max\_prio2$ with $size\_max\_prio2$. Furthermore, the size of the bitmaps used at 1-hop and 2-hop respectively depends on network topology.\\% We introduce the fields $size\_bitmap1$ and $size\_bitmap2$ to contain these sizes.\\

\subsubsection{Processing}\label{subsection-OSERENA-Rules}
%$ $\\
%\noindent \textbf{Rule R'0:} Any node $u$ computes  $max\_prio1(u)$, (respectively $max\_prio2(u)$, respectively $max\_prio3(u)$), only after it has received $priority(v)$, (respectively $max\_prio1(v)$, respectively $max\_prio2(v)$), from any 1-hop neighbor $v$.\\
\noindent With the optimization, Rules R1 and R2 become:\\

\noindent \textbf{Rule R'1:} Any node $u$ colors itself if and only if:\\ $priority(u) = max (max\_prio1(u), max\_prio2(u),$\\
\hspace*{60pt} $max\_prio3(u))$. $(eq.1)$\\

%\vspace*{3pt}
\noindent \textbf{Rule R'2:} When a node $u$ selects its color, it selects the smallest color unused in  $color\_bitmap1(u)$ $\cup$ $color\_bitmap2(u)$ $\cup$ $color\_bitmap3(u)$.\\

%\vspace*{3pt}
\noindent Notice that this color should also not be used by heard nodes (nodes with which there is no symmetric link). This, in order to avoid color conflicts.\\
\noindent The aim of rules R3 and R4 is to improve convergence time.
Although the $Color$ message does not contain the whole list of colored 2-hop neighbors, a node processing this message can deduce the recently colored nodes and stores them in a local data structure denoted $implicit\_node\_colored\_list$ whose size is equal to $implicit\_node\_colored\_size$. As we will see later, the storage of this list decreases the coloring time. To build this list, any node $u$ proceeds as follows:\\

\noindent \textbf{Rule R3:} When a node $u$ receives the $Color$ message from any neighbor node $v$ it compares the current value of $max\_prio1(v)$ (respectively $max\_prio2(v)$) with the previous one sent by $v$, denoted $previous\_max\_prio1(v)$ (respectively $previous\_max\_prio2(v)$). Any priority value of $previous\_max\_prio1(v)$ (respectively $previous\_max\_prio2(v)$) higher than
the highest value of $max\_prio1(v)$ (respectively $max\_prio2(v)$) corresponds to a recently colored node. This node is then inserted in the set $implict\_node\_colored\_list$.\\

\noindent \textbf{Rule R4:} When a node computes $max\_prio1$, $max\_prio2$ and $max\_prio3$ from the values received in the $Color$ messages, it proceeds as follows: %discards any value corresponding to an already colored node, provided that:
\begin{itemize}
\item in the computation of $max\_prio1$, it discards any priority value corresponding to an already colored node (that is a node that belongs to the list $implicit\_node\_colored\_list$).
\item in the computation of $max\_prio2$, it discards for any sender $v$, any priority value $p$ corresponding to an already colored node received in $max\_prio1(v)$ if and only if:
\begin{enumerate}
\item either $p$ is the highest priority in $max\_prio1(v)$,
\item or $p$ is the second highest priority in $max\_prio1(v)$ and (the third or fourth highest priority in $max\_prio1(v)$ is equal to $\varnothing$),
\item or $p$ is the third highest priority in $max\_prio1(v)$ and (the fourth highest priority in $max\_prio1(v)$ is equal to $\varnothing$).
\end{enumerate}
\item in the computation of $max\_prio3$, it discards for any neighbor $v$, any priority value corresponding to an already colored node received in $max\_prio2(v)$ if and only if it is the highest or the second highest priority in $max\_prio2(v)$.\\
\end{itemize}
The motivation of this rule is that a node $u$ can receive information about a node $v$ from a neighbor $w$ such that the distance between $v$ and $w$ is greater than the distance between $v$ and $u$. Consequently, node $w$ can send node $v$ as an uncolored node while $u$ knows that this node is colored. In such a case, if $u$ considers this information from $v$ and it sends it to its neighbors, coloring will be delayed because the node to be colored after $w$ will think this latter is uncolored and so does not color itself. However, using the list $implicit\_node\_colored\_list$, $u$ will discard the node $w$ and does not propagate an out-of-date information which helps to speed up the coloring convergence time. However, not any colored node can be discarded from $max\_prio1$ or $max\_prio2$.
Indeed, let us consider a node $u$ that discards the 4 values sent in $max\_prio1(v)$. In some configurations, $u$ might think it is the node having the highest priority among its 3 hop neighbors, although $v$ would have sent a priority higher than $priority(u)$ if it could send more than 4 values in a $max\_prio1$ it sends. That is why, we adopted the Rule 4. A node can be discarded from $max\_prio1$ or $max\_prio2$ if there is still at least one priority that may be equal to $\varnothing$.\\

%Intuitively, the priority of any colored node received in $max\_prio1(v)$ can be ignored if and only if there is at least one priority remaining in $max\_prio1(v)$, this remaining priority may be equal to $\varnothing$.\\

\noindent Rule R5 is related to the termination rule of the coloring algorithm.\\

\noindent \textbf{Rule R5:} Any node $u$ stops sending its $Color$ message as soon as it is colored, $max\_prio1(u)=\varnothing$ and it has received from all its 1-hop neighbors $v$ a $Color$ message with $max\_prio1(v)=max\_prio2(v)=\varnothing$.\\

\noindent Rule R6 has been introduced to tolerate message losses and link failures. \\

\noindent \textbf{Rule R6:} If at a round $r>1$ of the coloring algorithm, any node $u$ does not receive a message from its 1-hop neighbor $v$, it uses the information received from $v$ at round $r-1$. After $n$ successive rounds, with $n\geq2$ without receiving a $Color$ message from $v$, $v$ is no longer considered as a 1-hop neighbor of node $u$.\\

\section{Properties of \OSERENA{} \label{sec:PropOserena}}
%======================================================
\subsection{Correctness of \OSERENA{} coloring \label{subsec:validity}}
%----------------------------------------------------------------
In this section we prove that in a wireless environment assuming hypothesis A0, A1 and A2, \OSERENA{} provides a valid 3-hop node coloring avoiding collisions. Furthermore, we prove that this algorithm ends when all nodes are colored.
%\noindent \begin{lemma}\label{lemmaMaxPriocompleteConflictingSet}
%\textit{In an ideal wireless environment, any node $u$ has collected the useful information about all nodes in $\mathcal{N}(u)$ at any round $r\ge4$.}
%\end{lemma}
%\noindent \proof The first message $Color$ any node $u$ sends at round $1$ contains its priority. At round $2$, all receivers can compute their $max\_prio1$ and sends it to their neighbors. At round $3$, any receiver can compute its $max\_prio2$ and sends it in its $Color$ message. Consequently, in the round $r\geq 4$, assuming an ideal environment, any node would have received $max\_prio2$ from all its neighbors and can then compute its $max\_prio3$. % The node having the highest priority will color itself.
% Notice that since nodes store and exchange a specific number of the highest priorities of uncolored 1, 2, or 3-hop neighbors, we have: (1)
% $max\_prio1(u)$ can contain $u$ and/or its 1-hop neighbors; (2) $max\_prio2(u)$ can contain $u$ and/or its 1 and/or 2-hop neighbors; (3) $max\_prio3(u)$ can contain $u$ and/or its 1 and/or 2-hop neighbors. By definition, the set $max\_prio1$, (respectively  $max\_prio2$, respectively $max\_prio3$) is computed taking into account the 1-hop (respectively 2-hop, respectively 3-hop) neighbors.  
% That is why we say that \OSERENA{} allows any node $u$ to know the useful information about nodes in $\mathcal{N}(u)$.
%\endproof
\noindent \begin{lemma}
\label{lemmaWhenColor}
\textit{With \OSERENA{}, any node $u$ colors itself if and only if it has the highest priority among all the uncolored nodes in $\mathcal{N}(u)$.}
\end{lemma}
\noindent \proof Let us show that if any node $u$ is coloring itself, then it has the highest priority among the uncolored nodes up to 3-hop. By Rule R'1, if $u$ is coloring itself then $priority(u)= max (max\_prio1(u), max\_prio2(u),$\\
\hspace*{90pt} $max\_prio3(u))$ $(eq.1)$.\\
From $(eq.1)$, we get $priority(u) \geq max(max\_prio1(u))$. Hence, no uncolored one-hop neighbor has a priority higher than $u$.\\
From $(eq.1)$, we get $priority(u) \geq max(max\_prio2(u))$. Hence, no uncolored two-hop neighbor has a priority higher than $u$, otherwise we would have the following contradiction:\\ $priority(u) \geq max(max\_prio2(u)) > priority(u)$.\\
From $(eq.1)$, we get $priority(u) \geq max\_prio3(u)$. Hence, no uncolored three-hop neighbor in $\mathcal{N}(u)$ has a priority higher than $u$, otherwise we would have the following contradiction: $priority(u) \geq max\_prio3(u) > priority(u)$.\\
Hence, node $u$ has the highest priority among the uncolored nodes in $\mathcal{N}(u)$.\\

\vspace*{-0.3cm}
Conversely, if node $u$ has the highest priority among its uncolored neighbors up to 3-hop, it means that:
\begin{itemize}
\item all its uncolored one-hop neighbors have a smaller priority. Hence, for any $v$ uncolored one-hop neighbor of $u$, we have $priority(v) < priority(u)$. Hence,\\
$max(max\_prio1(u))=$\\
\indent $max_{\ v \in 1hop(u)} (priority(v)\ for\ v\ uncolored\ ) < priority(u)$;
\item all its uncolored two-hop neighbors have a smaller priority. Let us consider the highest priority in $max\_prio2(u)$. It denotes the highest priority of an uncolored node $w$ that is one-hop neighbor of $v$, itself one-hop neighbor of $u$. Consequently, we have the following cases:
\begin{itemize}
\item node $w$ is the node $u$ itself and has priority $priority(u)$;
\item node $w$ is a one-hop or two-hop neighbor of node $u$. In which case, we have by assumption: $priority(w)< priority(u)$.
\end{itemize}
Hence, $max(max\_prio2(u)) = priority(u)$.
\item and all its uncolored three-hop neighbors in $\mathcal{N}(u)$ have a smaller priority.
By definition, $max\_prio3(u)$ is the maximum priority of uncolored nodes $q$ that are one-hop neighbors of $w$, itself one-hop neighbor of $v$, one-hop neighbor of $u$. Consequently, we have the following cases:
\begin{itemize}
\item node $q$ is the node $u$ itself and has priority $priority(u)$;
\item node $q$ is a one-hop, two-hop or three-hop neighbor of node $u$. In which case, we have by assumption: $priority(q)< priority(u)$.
\end{itemize}
Hence, $max\_prio3(u) = priority(u)$.
\end{itemize}
Finally, $priority(u)= max (max\_prio1(u), max\_prio2(u),$\\
\hspace*{100pt} $max\_prio3(u))$.\\
Hence, node $u$ is coloring itself with \OSERENA{}.
%\noindent Conversely, let us assume $(eq.1)$: $priority(u)= max (max2\_prio1(u), max2\_prio2(u), max\_prio3(u))$. We now show than $u$ can color itself.\\
\endproof

\noindent \begin{lemma}\label{lemma-whichcolor}
\textit{With \OSERENA{}, when node $u$ colors itself, it knows all the colors taken in $\mathcal{N}(u)$ with a higher priority.}
\end{lemma}
\noindent \proof
%As shown in Lemma~\ref{lemmaMaxPriocompleteConflictingSet}, t
The exchange of $Color$ messages allows any node $u$ to know any uncolored node in $\mathcal{N}(u)$ having a higher priority than itself. Node $u$ also knows the colors of already colored nodes in $\mathcal{N}(u)$ by means of $bitmap1$, $bitmap2$ and $bitmap3$. Thus, when $u$ colors itself, it takes the smallest color unused in these bitmaps, and hence unused in $\mathcal{N}(u)$.
\endproof

%\noindent \begin{lemma}
%\textit{A node $u$ has the highest priority in $\mathcal{N}(u)$ if and only if $priority(u)=max (max\_prio1(u), max\_prio2(u), max\_prio3(u)$}.
%\end{lemma}
%\noindent \proof Let us assume any node $u$ having the highest priority in $\mathcal{N}(u)$.
%\endproof

\noindent \begin{lemma}
\textit{\OSERENA{} coloring ends when all nodes are colored.}
\end{lemma}
\noindent \proof If $u$ is colored and $max\_prio1(u)=\varnothing$, then node $u$ and all its one-hop neighbors are colored. Moreover, if node $u$ receives a $Color$ message from any one-hop neighbor $v$ with $max\_prio1(v)=max\_prio2(v)=\varnothing$, it means that all the one-hop neighbors of $v$ and all the one-hop neighbors of its one-hop neighbors are already colored. Hence, all nodes up to three-hop from $u$ and belonging to $\mathcal{N}(u)$ are colored. The coloring algorithm ends when node $u$ as well as all its 1-hop, 2-hop and 3-hop neighbors are colored.
\endproof

\noindent\begin{lemma}
\textit{In a wireless network meeting assumptions A0, A1 and A2 and in the absence of message loss and node failure, all nodes color themselves with \OSERENA{} and stop sending their $Color$ message.}
\end{lemma}
\noindent \proof
Let us consider any node $u$. The nodes in $\mathcal{N}(u)$ color themselves according to their priority. As soon as $u$ becomes the uncolored node with the highest priority, it colors itself according to rules R'1 and R'2.
According to rule R5, as soon as $u$ is colored and $max\_prio1(u)=\varnothing$, then node $u$ and all its one-hop neighbors are colored. Moreover, if node $u$ receives a $Color$ message from any one-hop neighbor $v$ with $max\_prio1(v)=max\_prio2(v)=\varnothing$, it means that all the one-hop neighbors of $v$ and all the one-hop neighbors of its one-hop neighbors are already colored. Hence, all nodes up to three-hop from $u$ and belonging to $\mathcal{N}(u)$ are colored. Hence, it is useless for $u$ to send its $Color$ message insofar as any information contained in its message is already known by its one-hop, two-hop and three-hop neighbors in $\mathcal{N}(u)$ and these nodes are already colored.
\endproof

\vspace*{0.3cm}
\noindent \begin{property}
\label{PropertyP1}
\textit{\OSERENA{} provides a valid 3-hop node coloring in any ideal wireless environment.}
\end{property}
\noindent \proof For three-hop coloring, for any node $u$, the set $\mathcal{N}(u)$ contains by definition all nodes up to 3-hop from $u$, assuming an ideal environment. %From Lemma~\ref{lemmaMaxPriocompleteConflictingSet}, any node $u$ has the information its needs to color itself. 
From Lemma~\ref{lemmaWhenColor}, with three-hop coloring, any node $u$ can color itself if and only if no uncolored node in $\mathcal{N}(u)$ has a priority higher than $u$.\\
According to rule R'1, priority of node $u$ meets $(eq.1)$. Moreover, since no two nodes have the same priority, we cannot have a simultaneous coloring of two nodes up to 3-hop away each other. According to Lemma~\ref{lemma-whichcolor}, when coloring itself, any node $u$ knows all the colors taken by nodes in its $\mathcal{N}(u)$, so it selects the smallest color according to rule R'2. Consequently, assuming an ideal wireless environment, no 2 nodes within 3-hop neighborhood from each other takes the same color. Which means that \OSERENA{} provides a valid coloring. With this coloring, nodes that belong to $\mathcal{N}(u)$ cannot create a collision with data sent by $u$ or an acknowledgement sent to $u$.
\endproof

\vspace*{0.3cm}
\noindent\begin{property}
\label{PropertyP2}
\textit{A failure to receive a $Color$ message from a one-hop neighbor induces an additional latency in network coloring and does not compromise the validity of coloring with \OSERENA{}.}
\end{property}
\noindent \proof
Deduced from rule R6.
\endproof

\vspace*{2pt}
\subsection{Equivalence of \OSERENA{} to a centralized algorithm\label{subsec:equivalence}}
%-----------------------------------------------------------------------------
In this section, we compare the behavior of \OSERENA{} with the well-known centralized First Fit 3-hop node coloring \cite{FirstFit}. More precisely, we compare the colors granted to nodes by both coloring algorithms.
With centralized First Fit 3-hop node coloring, nodes are sorted according to their priority and are colored in that order. Any node $u$ receives the smallest unused color in $\mathcal{N}(u)$.\\

\begin{lemma}
\label{lemma:RelativeOrder}
For any node $u$, for any given priority assignment, nodes $\in \mathcal{N}(u)$ color themselves in the same order with \OSERENA{} and First Fit.
\end{lemma}
\proof Let us consider any node $u$ that is coloring itself in \OSERENA{}, we have:
\begin{itemize}
\item any node $v \in \mathcal{N}(u)$ such that $priority(v) > priority(u)$ is already colored in \OSERENA{}, otherwise $u$ could not color itself now;
\item any node $v \in \mathcal{N}(u)$ such that $priority(v) < priority(u)$ is not colored in \OSERENA{}, because it is constrained by node $u$ that is not yet colored.
\end{itemize}
Hence, in $\mathcal{N}(u)$ the coloring order in \OSERENA{} is compliant with the priority order that is by definition followed by First Fit. In conclusion, both coloring algorithms follow the priority order to color nodes in a given neighborhood $\mathcal{N}(u)$.
\endproof

\vspace*{3pt}
\noindent \begin{property}
\label{PropertyP3}
\textit{For any topology, \OSERENA{} provides the same coloring as a centralized First Fit 3-hop node coloring algorithm using the same priority assignment.}
\end{property}
\proof For any topology, for any node $u$ in this topology, the color of $u$ is determined by the colors already used in $\mathcal{N}(u)$ when $u$ colors itself. According to Lemma \ref{lemma:RelativeOrder}, all nodes in $\mathcal{N}(u)$ color themselves in the same order with \OSERENA{} and First Fit. Let $u_1$ be the first node that colors itself in $\mathcal{N}(u)$. It takes the smallest available color in $\mathcal{N}(u_1)$. Let $u_2$ be the first node that colors itself in $\mathcal{N}(u_1)$, and so on. After a finite number of iterations (at most equal to the number of nodes in the topology), we get a node $u_{k+1}$ the first node that colors itself in $\mathcal{N}(u_k)$ and has colored itself without being constrained by any other node in \OSERENA{}: $u_{k+1}$ has the highest priority in $\mathcal{N}(u_{k+1})$. This node takes the color 0 in \OSERENA{}. With First Fit, since no node in $\mathcal{N}(u_{k+1})$ is already colored, $u_{k+1}$ takes color 0. Nodes in $\mathcal{N}(u_{k+1})$ with a priority higher than or equal to $priority(u_k)$ are colored according to their priority order with \OSERENA{} and First Fit. Consequently, they receive the same colors. We apply the same reasoning to node $u_k$ and nodes in $\mathcal{N}(u_{k})$ with a priority higher than or equal to $priority(u_{k-1})$, going back up to node $u_1$ and finally node $u$ that receives the same color with \OSERENA{} and First Fit, because the same colors are already assigned in $\mathcal{N}(u)$.
\endproof

\subsection{Reduced overhead\label{subsec:ReducedOverhead}}
%-------------------------------------------------
In this section, we show how \OSERENA{} reduces the overhead both in terms of 1) bandwidth by reducing message number and message size and 2) node storage by decreasing the size of data maintained at each node.\\

%\subsubsection{No maintenance of the 2-hop neighborhood}
%$ $\\
\OSERENA{} does not require to send or to maintain the 2-hop neighborhood of a node, as shown in Section~\ref{subsubsec:ColorMsg}. The use of $max\_prio1$ and $max\_prio2$ reduces the size of $Color$ messages exchanged between neighbors. We now show how to determine the optimal size of $max\_prio1$ and $max\_prio2$.

\vspace*{5pt}
\subsubsection{Message size}
%----------------------------

Assuming the near optimal size of $max\_prio1$ and $max\_prio2$ determined later on (see Lemma \ref{nearoptimalsize}), we can compute the maximum size of the message $Color$ exchanged between neighbor nodes.

\noindent \begin{property}
\label{PropertyP4}
\textit{With the setting $Size\_max\_prio1 = 4$ and $Size\_max\_prio2 =3$, \OSERENA{} uses a $Color$ message whose size is at most $8 \cdot (size\_address + size\_prio) + size\_color+ size\_bitmap1 + size\_bitmap2$ bytes.}
\end{property}
\noindent \proof This is deduced from the $Color$ message format, where the 8 factor comes the maximum size of $priority + max\_prio1 + max\_prio2$. %occupy respectively twelve bytes and nine bytes (three bytes per priority including one byte for the $prio$ component and two bytes for the $address$ component).
\endproof
%$ $\\

\vspace*{5pt}
\subsubsection{Constraints for the computation of $max\_prio1$ and $max\_prio2$ sizes}
%----------------------------
%$ $\\
We first notice that the reduction of message size must not imply a higher number of rounds to color the network. Hence, the optimal size of $max\_prio1$ and $max\_prio2$ is a trade-off  between bandwidth consumption and convergence time of the coloring algorithm.\\

%Objective: show how it was not easy to find the relevant size of the fields maxprio1 and maxprio2 minimizing the
%number of rounds needed to color the network: (tradeoff between bandwidth and number of rounds)
The simplest solution would be to maintain only one priority for $max\_prio1$ and $max\_prio2$. However, this solution does not allow to remove already colored nodes in the computation of $max\_prio1$, $max\_prio2$ and $max\_prio3$. Hence, a coloring that is much slower than \SERENA{}.
% as illustrated by scenario1
%. In scenario1, XXX nodes uniformly distributed with a density of XXX need XXX rounds to color themselves instead of XXX rounds with \SERENA{}. 
That is why, several priorities are maintained in $max\_prio1$ and $max\_prio2$. The question is how many? To be able to discard one value corresponding to an already colored node and sent by neighbor $v$ in $max\_prio2(v)$ implies that $v$ sends at least 2 values in $max\_prio2(v)$. To be able to compute its 2 highest values in  $max\_prio2(v)$ and discard one value, node $u$ must receive at least 3 values in $max\_prio1(u)$. Hence, the minimum sizes are $Size\_max\_prio1 = 3$ and $Size\_max\_prio2 = 2$.\\
 %if we want to accelerate the coloring by removing already colored nodes in the computation of $max\_prio1$, $max\_prio2$ and $max\_prio3$, we can get a scenario1 to show why the version of 'Future internet' does not work
% if we want to accelerate the coloring by removing colored nodes from $maxprio1$, $maxprio2$ and $maxprio3$
%\item transition: show that the node should not ignore any address in $maxprio2$ even if it corresponds to a colored node because when it colors itself it should have an information
 %\item

Unfortunately, we can still exhibit scenarios %, like scenario2, 
with this minimum setting, where only the first address in $max\_prio2$ is discarded if already colored, producing a number of rounds higher than \SERENA{}. 
In simulations, we identified a scenario with
$100$ nodes uniformly distributed with a density of $20$ need $175$ rounds to color themselves with \OSERENA{} instead of $134$ rounds with \SERENA{}. 
That is why, we select $Size\_max\_prio1 = 4$ and $Size\_max\_prio2 =3$.\\

 \subsubsection{Computation of the optimal size of $max\_prio1$ and $max\_prio2$}
 %--------------------------------------------------------------------------------
 %$ $\\
In this section, we assume the unit disk graph model of section~\ref{sec:simplify} 
(including \emph{Assumption A3}).

 Recall that in \OSERENA{}, any node $u$ receiving a
$Color$ message from one neighbor $v$, can have fresher information than $v$. That is, $v$ believes that node $w$ is not yet colored and so, keeps it in $max\_prio1(v)$
or $max\_prio2(v)$ it sends, whereas $u$ knows that $w$ is already colored because it is closer to $w$ than $v$. In such a case, \OSERENA{} allows $u$ to ignore $w$ when computing
$max\_prio1(u)$ by usage of the list $implicit\_node\_colored\_list$ as explained in section~\ref{subsection-OSERENA-Rules}. However, to keep the correctness of the algorithm, the node $u$ cannot always ignore the colored node $w$ when computing $max\_prio2(u)$ and $max\_prio3(u)$ (see the coloring rule R3). If
node $u$ that is the next node to be colored after $w$, is not allowed to ignore $w$ already colored, $u$ will not color itself and will wait until node $v$ removes node $w$. Hence, node $u$ in \OSERENA{} colors itself later than it would do in \SERENA{}.\\

\vspace*{-3pt}
 \begin{lemma}
 \label{nearoptimalsize}
 With the setting $Size\_max\_prio1 = 4$ and $Size\_max\_prio2 = 3$ and rules R3 and R4, \OSERENA{} colors any node $u$ in the same round as \SERENA{}, except when three nodes two-hop away from $u$, but 4-hop away from each other are coloring simultaneously just before $u$.
 \end{lemma}
 \proof We first identify this scenario and then compute its probability in the next section. When three nodes two-hop away from $u$, but 4-hop away from each other are coloring simultaneously just before $u$, node $u$ is not allowed by rule R3 to discard the three of them in the received $max\_prio2(v)$, hence the coloring of node $u$ is delayed.
We can show that this scenario is the only one that will delay $u$ coloring. On the one hand, two one-hop neighbors of $u$ are not allowed to color simultaneously, because they are at most two-hop away. On the other hand, a one-hop and a two-hop neighbor of $u$ are not allowed to color simultaneously, because they are at most three-hop away. It results that the only case of simultaneous colorings in $\mathcal{N}(u)$ involves nodes that are 2-hop away from $u$ and 4-hop away from each other.
 \endproof
%     scenario2: to show that {if maxprio1Size = 3 and maxprio2Size =2 and if the
%    node does not take into account only the first address in maxPrio2 if it is a colored node} produces high number of rounds
 %    \item (potential question: comparison of gain and loss
 %   between {maxprio1Size=3+maxprio2Size=2 and maxprio1Size=4+maxprio2Size=3} =>because the number of rounds {if maxprio1Size = 3 and maxprio2Size =2 and if the node does not take
 %   into account only the first address in maxPrio2 if it is a colored node} is considerably higher than \SERENA{} (give an example), we studied the case
%    (maxprio1Size=4+maxprio2Size=3)....
\vspace*{+5pt}
\begin{lemma}
The setting $Size\_max\_prio1 = 5$ and $Size\_max\_prio2 =4$ provides the same number of rounds as \SERENA{}.
\end{lemma}
\proof
With the setting, $Size\_max\_prio1 = 5$ and $Size\_max\_prio2 =4$, it is no longer possible to have a bad scenario where four nodes two-hop away from $u$, but 4-hop away from each other are coloring simultaneously. We prove it by contradiction. Let $u$ be any node. We assume that the four nodes $v_1$, $v_2$, $v_3$ and $v_4$ that are 4-hop away from each other and 2-hop away from $u$ are coloring themselves simultaneously. We notice that the distance between these four nodes is maximized when they belong to the circle centered at $u$ and of radius $2R$ and are diametrally opposed. We can compute the distance of two adjacent points denoted $v_1$ and $v_2$, we then have $d(v_1, v_2)^2= d(v_1,u)^2+d(u,v_2)^2=4R^2+4R^2=8R^2$. Hence $d(v_1,v_2)=2\sqrt{2}R<3R$: this contradicts our assumption.
\endproof

\vspace*{3pt}
That is why in the following of this research report, we take $Size\_max\_prio1 = 4$ and $Size\_max\_prio2 = 3$ leading to a smaller bandwidth use.

\subsection{Convergence time}
%---------------------------------
As shown in the previous section, the selected setting of the size of $max\_prio1$ and $max\_prio2$ provides the same number of rounds as \SERENA{}, except when the bad scenario occurs. In the bad scenario, the coloring of a node is delayed in \OSERENA{}.
Notice that even in this case, the total number of rounds required by \OSERENA{} can still be equal to the total number of rounds required by \SERENA{}. The occurrence of the bad scenario is a necessary but not sufficient condition to increase the number of rounds with \OSERENA{}.
To conclude, the scenario where one \OSERENA{} node $u$ is colored with a delay compared to \SERENA{} happens if the following events occur:
\begin{itemize}
\item \emph{$E_1$:} $\exists$ $v_1$, $v_2$ and $v_3$, three nodes that are 2-hop away from $u$
and 4-hop away from each other.
\item \emph{$E_2$:} these three nodes $v_1$, $v_2$ and $v_3$ have a priority higher than $u$.
\item \emph{$E_3$:} $v_1$, $v_2$ and $v_3$ are colored simultaneously.\\
 \end {itemize}
%ZZZZZZZZZZZZZZZZZZZZZZZZZZZZZZZZZZZZZZZZZZZZZZZZZZZZZZZZZZZZZZZZZZZZZZZZZZZ

%\subsubsection{Evaluation of the probability of the existence of a convergence time difference between \SERENA{} and \OSERENA{}}
%---------------------------------------------------------------------------------------
%$ $\\
We assume the unit disk graph model of section~\ref{sec:simplify},
including \emph{Assumption A3}.
We adopt the following notations. Let $d(u,v)$ denote the euclidian distance between nodes $u$ and $v$. Let $P$ denote the probability that the bad scenario occurs. We want to estimate an upper bound of this probability.
Let $P_i$ denote the probability that the event $E_i$ occurs, with $i \in[1,3]$. We have:
\vspace*{-5pt}
\begin{displaymath}
\vspace*{-5pt}
P \leq \prod _{i=1}^3 P_i.
\end{displaymath}

\begin{figure}[!ht]
\centering
\includegraphics[width=2.7in]{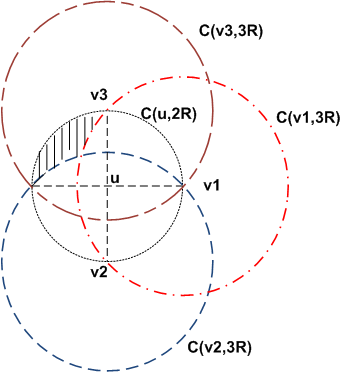}
\caption{Possible zone for node $v_3$.}%A figure illustrating the scenario described by $Z_2,Z_1,E_3$.}
\label{fig:v1v2v3}
\end{figure}

\vspace*{-5pt}
For any node $u$, let $\mathcal{D}(u,R)$, (respectively $\mathcal{C}(u,R)$), denote the disk (respectively the circle) centered at $u$ of radius $R$. Let $A \setminus B$ denote the set containing exactly the elements of $A$ but not those of $B$.\\
The computation of upper bounds of probabilities $P_1$ and $P_2$ is done geometrically. On Figure ~\ref{fig:v1v2v3}, a bound of $P_1$ corresponds to the probability for $v_3$ to belong to the hatched area.%, whereas $P_3$ corresponds to the probability to belong to the dotted area.

\subsubsection{Estimation of an upper bound of $P_1$}
\begin{figure}[!ht]
\centering
\includegraphics[width=2.5in]{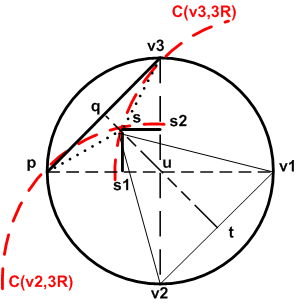}
\caption{A figure illustrating the computation of $P_1$.}
\label{fig:probaComputing}
\end{figure}

%---------------------------------------------------
$ $\\
The computation of the probability $P_1$ is illustrated in figure~\ref{fig:probaComputing}.
% Let $\overline A$ denote the complement of A.\\
Nodes $v_i$, for $i \in[1,3]$, should belong to $\mathcal{D}(u,2R) \setminus \mathcal{D}(u,R)$ and should be at a distance belonging to $(3R,4R]$ from each other. To maximize the number of possible nodes $v_3$, we take $v_1 \in \mathcal{C}(u,2R)$. The choice of $v_1$ done, we increase the number of possible nodes $v_3$ by taking $v_2 \in  \mathcal{C}(u,2R) \cap \mathcal{C}(v_1,3R)$, approximating $3R+\varepsilon$ by $3R$. We make $v_1$ and $v_2$ closer increasing again the possibilities for $v_3$ by transforming the triangle $(v_1, u, v_2)$ in a right triangle. We then have $d(v_1,v_2)=2R \sqrt{2}$ computed as the hypotenuse in the triangle $(u,v_1,v_2)$. We now select $v_3$ that belongs to $\mathcal{D}(u,2R) \setminus \mathcal{D}(u,R) \setminus \mathcal{D}(v_1,2R \sqrt{2}) \setminus \mathcal{D}(v_2,2R \sqrt{2})$, corresponding to the hatched area depicted in Figure~\ref{fig:v1v2v3}. We compute $S_P$ the surface of this area.
$S_P \leq S_D - 2 S_T - S_C$, where $S_D$ is the surface of the disk quarter $\mathcal{D}(u,2R)$, $S_T$ is the surface of the triangle formed by $s$, $u$ and $v_3$ and $S_C$ the surface of the square $(s,s_1,u,s_2)$ whose diagonal is $y$ (see figure \ref{fig:v1v2v3}). We first compute $d(u,q)$ in the right triangle $(u, q, v_3)$. We get $2 R^2 +d(u,q)^2=2^2 R^2$. Hence, $d(u,q)=R\sqrt{2}$. In the isosceles triangle $(v_1,v_2,s)$, we compute $d(q,t$. We have: $(d(s,t))^2+2 R^2=8R^2$. Since $d(s,t)=d(s,u)+d(u,q)$, we get $d(s,u)=d(s,t)-d(u,q)=(\sqrt{6}-\sqrt{2})R$. We then get:\\
$S_D= \Pi R^2$ and $S_C=y^2/2=(4-2\sqrt{3})R^2$.\\
$S_T = (\sqrt{4-2\sqrt{3}}-2+\sqrt{3})R^2$.\\
We deduce $S_P = (\Pi-2\sqrt{4-2\sqrt{3}})R^2$.\\
Hence, $P_1= \frac{number\ of\ favorable\ cases}{number\ of\ possible\ cases}= \frac{S_P}{4 \Pi R^2}$.\\
Finally, we get $P_1= \frac{1}{4}-\frac{\sqrt{4-2\sqrt{3}}}{2\Pi}$.

\vspace*{3pt}
\subsubsection{Estimation of an upper bound of $P_2$}
%---------------------------------------------------
%$ $\\
For any node $u$, let us compute $P_2$ the probability of event $E_2$: there exists three nodes two-hop away from $u$ with a priority higher than $u$. This event $E_2$ can be considered as the intersection of two events $E_{21}$ and $E_{22}$, where $E_{21}$ means that there exists three nodes in $\mathcal{D}(u,2R)$ with a priority higher than $u$. Event $E_{22}$ means that three nodes in $\mathcal{D}(u,2R)$ do not belong to $\mathcal{D}(u,R)$. We do not have event $E_21$ if and only if in $\mathcal{D}(u,2R)$, 1) $u$ has the highest probability, or 2) $u$ has the second highest probability or 3) $u$ has the third highest probability. Let $M$ denote the number of nodes that are exactly one-hop away from $u$. The average number of nodes in $\mathcal{D}(u,2R)$ is equal to $4M +1$. We compute $P_{21}$ the probability of event $E_{21}$. We have $P_{21}=1-\frac{3}{4M+1}$.\\ 
We can now compute $P_{22}$ the probability of event $E_{22}$. We get $P_{22}$=probability that none of these three nodes in $\mathcal{D}(u,2R)$ belong to $\mathcal{D}(u,R)$. Since the nodes are independent, we get $P_{22}=(1-\frac{\Pi R^2}{4\Pi R^2})^3=(3/4)^3$. Since events $E_{21}$ and events $E_{22}$ are independent, we get $P_2=P_{21} \cdot P_{22}$, leading to $P_2= \frac{27}{64}(1-\frac{3}{4M+1})$.
\\
% and $M=\frac{3\Pi R^2}{N} density$, where $N$ is the total number of nodes.\\

\subsubsection{Estimation of an upper bound of $P_3$}
%---------------------------------------------------
$ $\\
For any node $u$, we select the last three nodes $v_1$, $v_2$ and $v_3$, two-hop away from $u$ that color themselves just before $u$. We want to compute $P_3$ the probability that event $E_3$ occurs that is: these three nodes color themselves simultaneously. 
We can bound $P_3$ by 1.\\

\subsubsection{Upper bound for $P$}\label{sec:upper-bound}
%--------------------------------
%$ $\\
\begin{property}
\label{P5}
The probability of occurrence of the bad scenario is upper bounded by $\frac{27}{64}(1-\frac{3}{4M+1})\cdot (\frac{1}{4}-\frac{\sqrt{4-2\sqrt{3}}}{2\Pi})$.
\end{property}
\proof Since $P \leq \prod _{i=1}^3 P_i$, we get $P \leq \frac{27}{64}(1-\frac{3}{4M+1})\cdot (\frac{1}{4}-\frac{\sqrt{4-2\sqrt{3}}}{2\Pi})$.
\endproof
~\\
Noticing that $(1-\frac{3}{4M+1}) \le 1$, a numeric evaluation of the bound yields: $P \le 0.0564$
% python -c 'from math import * ; print 27/64.0 * (1/4.0 - sqrt(4-2*sqrt(3)) / (2*pi))'

%Notice that we made statistics on the occurrence number of this bad scenario in random topologies where nodes are uniformly distributed in the network area. Statistics results show that $P \leq 3\cdot10^{-6}$ XXX.

\section{Performance evaluation by simulation \label{PerfSimul}}
%==============================================
We now evaluate the performance of \OSERENA{} by simulation for various WSNs.

\subsection{Simulation modules and parameters}
%---------------------------------
We consider various wireless network configurations, with the unit disk model, where the number of nodes varies from $50$ to $200$ and the average number of neighbors per node, called density, varies from $8$ to $45$.  We check the connectivity of all the topologies generated by our random topology generator. %Furthermore, the average number of neighbors per node is equal to $\Pi \cdot R^2 \cdot density$.

\noindent Three modules are simulated:
\begin{itemize}
\item \textit{The Neighborhood Discovery Module} in charge of detecting the creation of new links, testing their symmetry and detecting their breakdown. This is done by means of periodic exchanges of $Hello$ messages. The $Hello$ message contains the list of addresses of heard/symmetric nodes.
\item \textit{The \OSERENA{} Module} in charge of coloring the wireless network, once topology is stabilized.
\item \textit{The \SERENA{} Module} used as a reference for a comparative performance evaluation.
\end{itemize}

%\noindent The parameters taken are the following. The MAC protocol is XXX\\

\noindent We evaluate the number of colors used, the number of rounds needed to color the whole network, the average number of $Color$ messages sent per node as well as the average size of these messages.
Each result is the average of $10$ to $50$ simulations.

%The goal of \OSERENA{} is to reduce data stored and exchanged at a small expense (rounds slightly larger than \SERENA{}). Consequently, it is interesting to show that the average message size in \OSERENA{} and the average  size of data stored are smaller than those required by \SERENA{}. As far as the message size, I'm afraid we can't show it is smaller because it depends on the rounds number and the number of message....

\subsection{Performance results of \OSERENA{} \label{subsec:OSERENAPerf}}
%---------------------------------------------
In this series of simulations, we fix the number of nodes in the interval $[50, 200]$ and vary the node density from $8$ to $45$. We evaluate the performance criteria of \OSERENA{} and then iterate on another number of nodes.\\

\subsubsection{Number of colors\label{subsubsec:colors}}
%---------------------------------
$ $\\
The main performance criterion of a coloring algorithm is the number of colors needed to color the whole network. This number depends on network topology. First, we want to evaluate the impact of node density and node number on the number of colors used by \OSERENA{}.
 
\begin{figure}[!htb]
\centering
\includegraphics[width=\figsize]{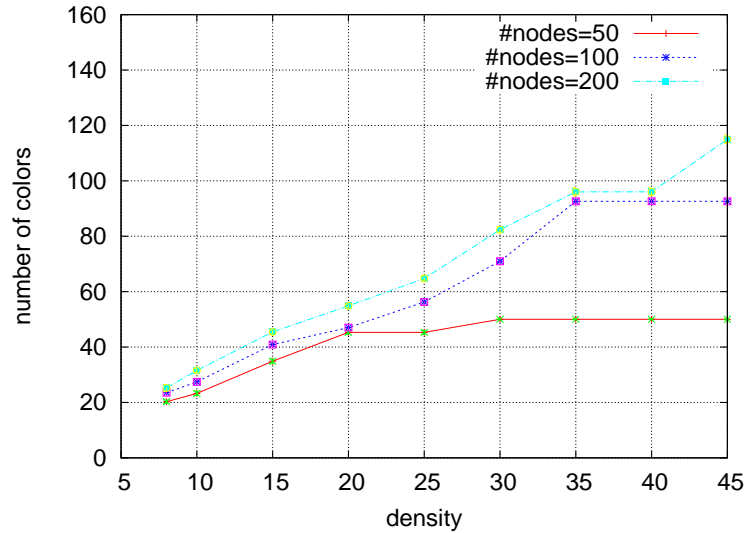}
\caption{Number of colors.}
\label{fig_Colors}
\end{figure}
The figure~\ref{fig_Colors} shows that the number of colors strongly depends
on the density of the graphs, and much less on the number of nodes. 
Intuitively, the reason is that the color selected by a node, depends only on
its $3$-hop neighborhood, hence is related to the number of the $3$-hop
neighbors (which is itself directly proportional to density). 

Furthermore, the size of the $3$-hop neighborhood is not related to the number of nodes
of the graph, hence this last parameter has less impact.
This occurs until the transmission range becomes too large and the $3$-hop neighborhood
includes the whole network (as shown in the figure for a number of nodes
$=$ 50 and for density $\ge$30, where increasing density for a fixed 
number of nodes is equivalent to increasing transmission range).

\subsubsection{Number of rounds\label{subsubsec:rounds}}
%---------------------------------
$ $\\
To measure the time complexity of \OSERENA{}, we evaluate the number of rounds needed to color the whole network. More precisely, what is the impact of node density and node number on the number of rounds? 

\begin{figure}[!htb]
\centering
\includegraphics[width=\figsize]{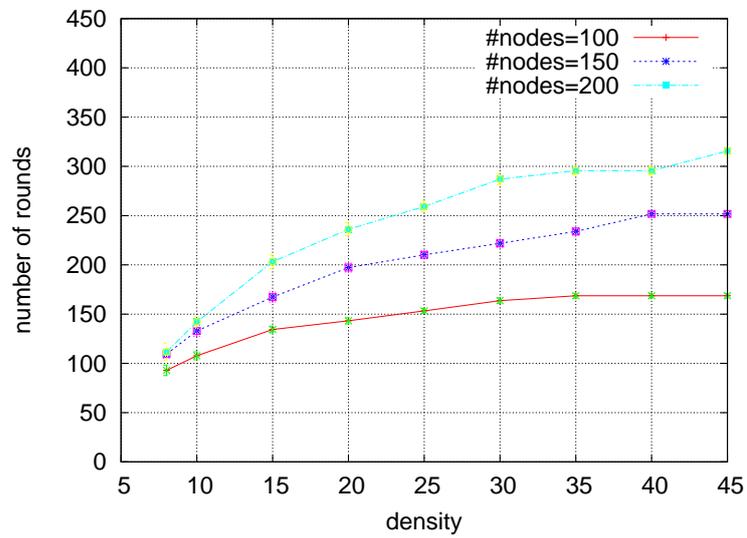}
\caption{Number of rounds.}
\label{fig_Rounds}
\end{figure}
In figure~\ref{fig_Rounds}, we observe that the number of rounds depends 
more on the number of nodes in the network than on density.

There is one natural explanation on the observation that the number of nodes 
has an impact on the number of rounds (and much less on the number of colors, see previous
section): in \OSERENA{}, every node $u$ must wait until all the nodes in its $\mathcal{N}(u)$ having a higher priority than itself color themselves. Recursively, each node in this set should do the same. 
This is likely to lead to waiting ``chains'',
and such chains are longer in larger networks.
It contributes to increase coloring delay.

%its $3$-hop
%neighborhood because of which it has to delay its coloring 
%Moreover, the more the density is 

\subsubsection{Number of messages sent per node\label{subsubsec:msg}}
%---------------------------------
$ $\\
To compute the overhead induced by \OSERENA{}, we first evaluate the average number of messages sent per node for various network configurations, pointing out the influence of node density and node number.

\begin{figure}[!htb]
\centering
\includegraphics[width=\figsize]{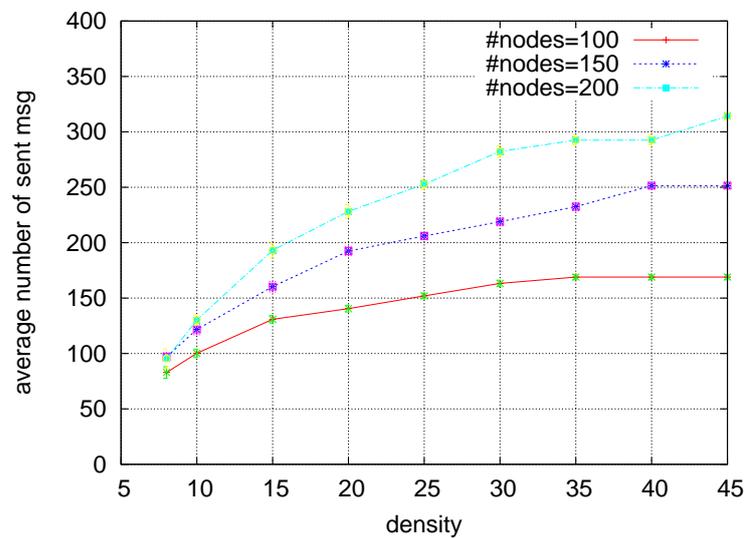}
\caption{Average number of messages sent per node.}
\label{fig_MsgNumber}
\end{figure}
As illustrated in figure~\ref{fig_MsgNumber}, the average number of messages
is close to the number of rounds (in figure~\ref{fig_Rounds}). 
This is expected since every node sends one message per round until a stopping 
condition is fulfilled (rule R5):
in the simulations, for most nodes, most of the time, rule R5 is not verified.

\subsubsection{Number of bytes sent per node\label{subsubsec:byte}}
%---------------------------------
$ $\\
Another expression of the message overhead is given by the average number of bytes sent per node for various network configurations. What is the impact of node density and node number on the number of bytes exchanged during the coloring?

\begin{figure}[!htb]
\centering
\includegraphics[width=\figsize]{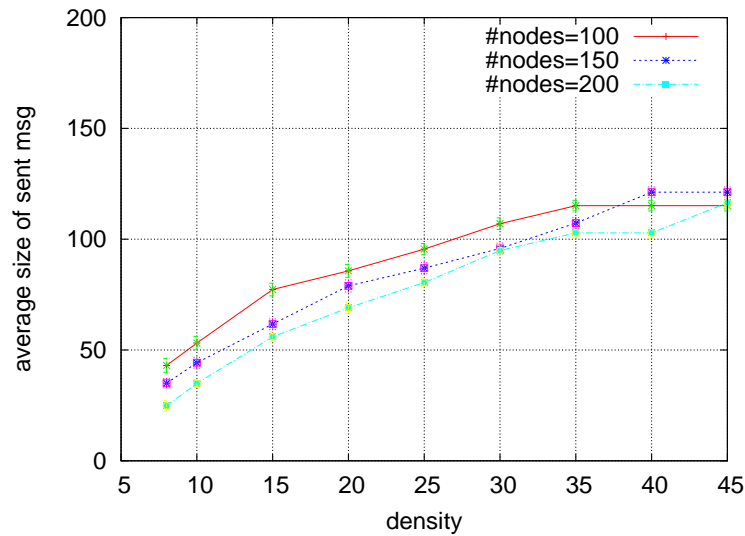}
\caption{Average number of bytes sent per node.}
\label{fig_ByteNumber}
\end{figure}
From the figure~\ref{fig_ByteNumber}, the number of nodes has barely 
noticeable impact on the average number of bytes sent per nodes, 
whereas density has a limited, but direct impact. 
This is a direct consequence of the structure of the $Color$ message,
which includes $2$ bitmaps of colors of the $1$-hop and $2$-hop
neighborhood, which increases linearly with density (e.g.  $2$ additional
bits in message, per additional color in the $3$-hop neighborhood).

\subsection{Comparison with \SERENA{}\label{subsec:comparison}}
%---------------------------------------------
We now compare the performances obtained by \OSERENA{} with those of \SERENA{}. \OSERENA{} ensures that nodes should get the same colors as with \SERENA{}. The open question is at which expense?
%XXX: Does the bad scenario occur frequently? Does-it systematically lead to an increased number of rounds?\\

\subsubsection{Number of colors\label{subsubsec:colorscomp}}
%---------------------------------
$ $\\
Simulation results are compliant with the expected behavior of \OSERENA{}: any node receives the same color with \SERENA{} and \OSERENA{}.\\

\subsubsection{Number of rounds\label{subsubsec:roundscomp}}
%---------------------------------
$ $\\
Simulation results show that even if the bad scenario occurs, \OSERENA{} needs
the same number of rounds as \SERENA{} in all the network topologies tested.

\begin{figure}[!htb]
\centering
\includegraphics[width=\figsize]{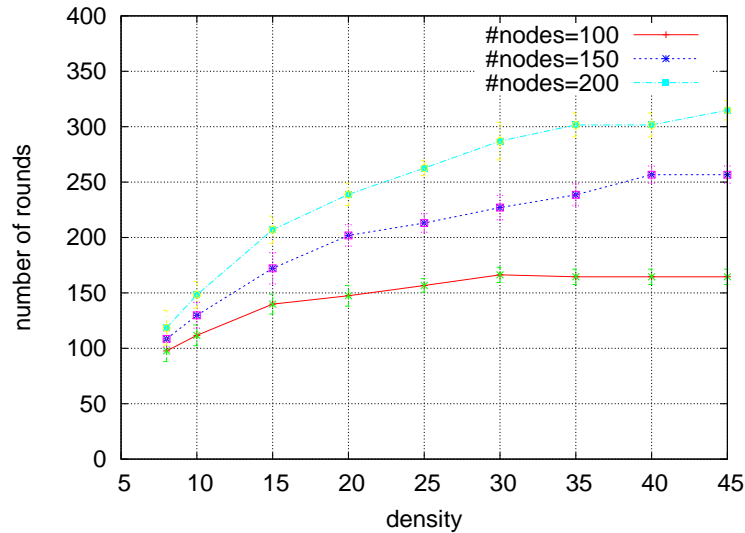}
\caption{Number of rounds for \SERENA{}}
\label{fig_ByteNumberSerena}
\end{figure}
Comparing figure~\ref{fig_ByteNumber} and figure~\ref{fig_ByteNumberSerena}
we observe that the number of rounds is equivalent. The reason is that the
event where \OSERENA{} requires more rounds than \SERENA{} on one node
has low probability (see section~\ref{sec:upper-bound}) ; and then,
the occurrence of one such event does not automatically increase the total
number of rounds for the coloring of the whole network.

\subsubsection{Number of messages sent per node\label{subsubsec:msgcomp}}
%---------------------------------
$ $\\
Simulation results show that the average number of messages sent per node is comparable with \SERENA{} and \OSERENA{} for various network configurations. This is a consequence of the previous result.

\begin{figure}[!ht]
\centering
\includegraphics[width=\figsize]{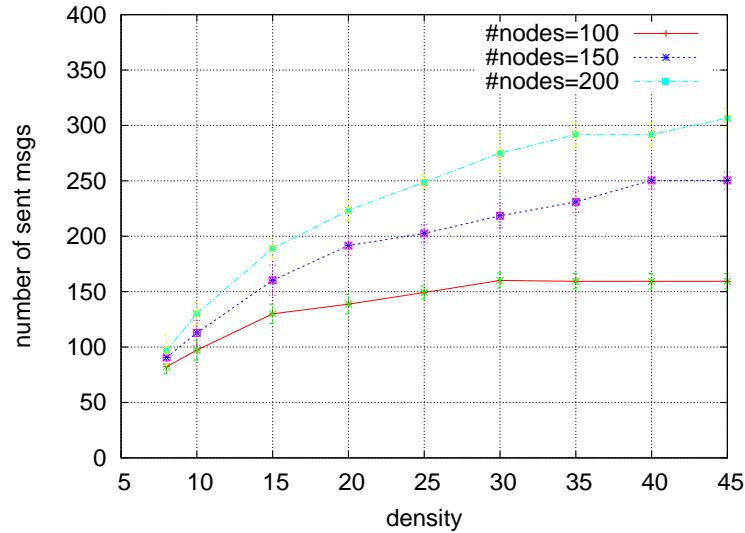}
\caption{Average number of messages sent per node with \SERENA{} and \OSERENA{}.}
\label{fig_MsgNumberComp}
\end{figure}

\subsubsection{Number of bytes sent per node\label{subsubsec:bytecomp}}
\begin{figure}[!ht]
\centering
\BlindReview{
% blind VERSION
\includegraphics[width=\figsize]{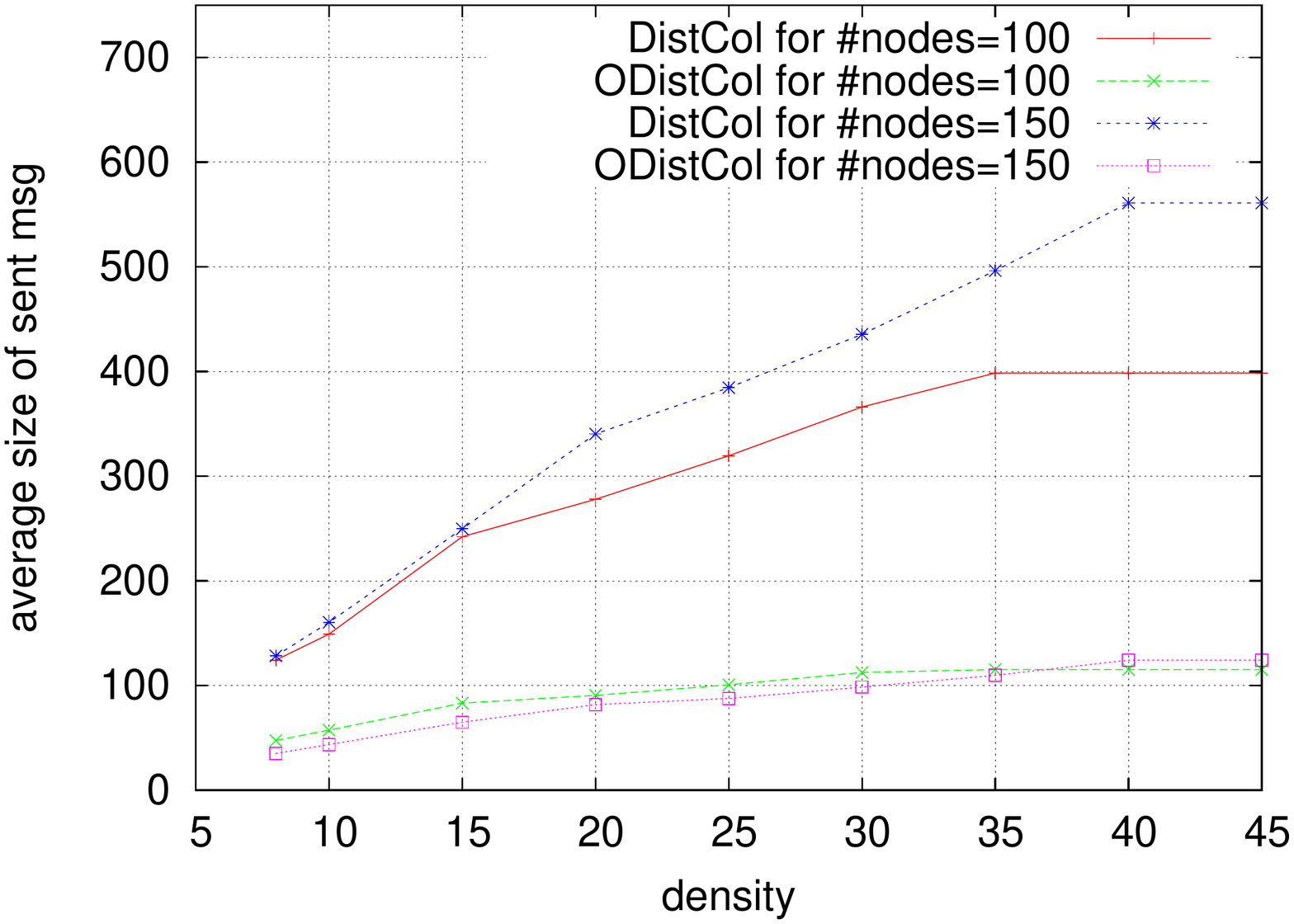}
}{
% normal VERSION
\includegraphics[width=\figsize]{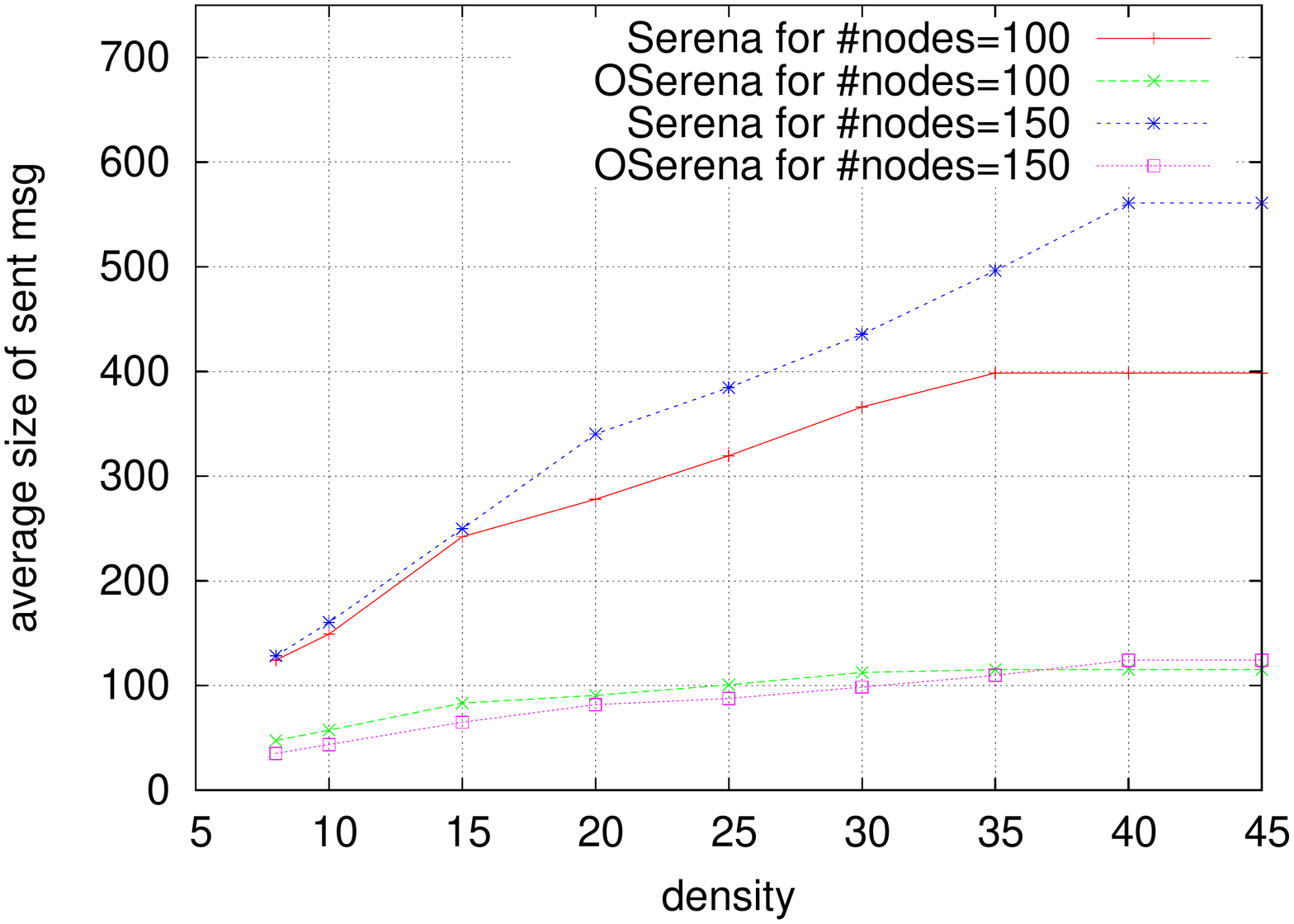}
}
\caption{Average number of bytes sent per node with \SERENA{} and \OSERENA{}.}
\label{fig_ByteNumberComp}
\end{figure}

%---------------------------------
$ $\\
Figure~\ref{fig_ByteNumberComp} depicts the average number of bytes sent per node with \SERENA{} and \OSERENA{}. It points out the benefit brought by the optimization of \OSERENA{}.

The figure~\ref{fig_ByteNumberComp} illustrates the major contribution of
\OSERENA{} compared \SERENA{}: the size of the $Color$ messages is much 
smaller.

In \SERENA{}, a $Color$ message includes information for each  node in
its entire $3$-hop neighborhood (address, priority, color): several bytes
per node in the $3$-hop neighborhood. In \OSERENA{}, only a
small fixed subset of priorities and addresses of these nodes are exchanged,
and only $2$ bits per color are required.

Notice that for wireless sensor networks based on 802.15.4, the maximum
packet size is $127$ bytes, hence \SERENA{} messages are problematic
even at the lowest density (and would have probably to be fragmented
in several packets), 
whereas on contrary, \OSERENA{} fits within this limit until high densities.

%\begin{figure*}[!t]
%\centerline{\subfloat[Case I]\includegraphics[width=\figsize]{subfigcase1}%
%\hfil
%\subfloat[Case II]{\includegraphics[width=\figsize]{subfigcase2}%
%\label{fig_second_case}}}
%\caption{Simulation results}
%\label{fig_sim}
%\end{figure*}
%
% Note that often IEEE papers with subfigures do not employ subfigure
% captions (using the optional argument to \subfloat), but instead will
% reference/describe all of them (a), (b), etc., within the main caption.

\section{Conclusion\label{Conclusion}}
%Dense WSNs are a real challenge for the design of bandwidth and energy efficient distributed protocols.
Coloring algorithms have been introduced in WSNs to allow sensor nodes to save energy and bandwidth. Collisions are avoided and nodes can sleep when they are neither sender nor receiver of the transmitted messages. However, their use in dense WSNs is possible only if they are optimized to support such networks. Indeed, the resource constrained nature of sensors combined with the possible high number of neighbors is a real challenge for the design of bandwidth and energy efficient protocols. That is why we have proposed \OSERENA{}, whose performance evaluation results confirm that the WSN is colored with the almost the same number of rounds and exactly the same number of colors as its unoptimized version, but with a message size that does not depend on network density. Consequently, \OSERENA{} enables considerable gains in bandwidth and energy consumption.

\end{document}